\documentclass[sigconf,authorversion,nonacm]{acmart}
\AtBeginDocument{%
  }

\usepackage{tabularx}
\usepackage{booktabs} 

\usepackage{amsmath} 

\usepackage{array}
\usepackage{multirow}
\usepackage{booktabs}

\usepackage{listings} 
\usepackage{xcolor}   

\lstdefinestyle{structuredText}{
    backgroundcolor=\color{white},   
    basicstyle=\ttfamily\small,      
    breaklines=true,                 
    captionpos=b,                    
    frame=single,                    
    keywordstyle=\color{blue},       
    commentstyle=\color{green},      
    stringstyle=\color{red},         
    numbers=left,                    
    numberstyle=\tiny\color{gray},   
    tabsize=4,                       
    showstringspaces=false           
}

\setcopyright{acmlicensed}
\copyrightyear{2018}
\acmYear{2018}
\acmDOI{XXXXXXX.XXXXXXX}
\acmConference[Conference acronym 'XX]{Make sure to enter the correct
  conference title from your rights confirmation email}{June 03--05,
  2018}{Woodstock, NY}
\acmISBN{978-1-4503-XXXX-X/2018/06}

\begin{document}

\title{AttackLLM: LLM-based Attack Pattern Generation for an Industrial Control System}

\author{Chuadhry Mujeeb Ahmed}
\email{mujeeb.ahmed@newcastle.ac.uk}
\affiliation{%
  \institution{Newcastle University}
  \country{UK}
}








\renewcommand{\shortauthors}{CM Ahmed}

\begin{abstract}
Malicious examples are crucial for evaluating the robustness of machine learning algorithms under attack, particularly in Industrial Control Systems (ICS). However, collecting normal and attack data in ICS environments is challenging due to the scarcity of testbeds and the high cost of human expertise. Existing datasets are often limited by the domain expertise of practitioners, making the process costly and inefficient. The lack of comprehensive attack pattern data poses a significant problem for developing robust anomaly detection methods. In this paper, we propose a novel approach that combines data-centric and design-centric methodologies to generate attack patterns using large language models (LLMs). Our results demonstrate that the attack patterns generated by LLMs not only surpass the quality and quantity of those created by human experts but also offer a scalable solution that does not rely on expensive testbeds or pre-existing attack examples. This multi-agent based approach presents a promising avenue for enhancing the security and resilience of ICS environments.
\end{abstract}




\received{20 February 2007}
\received[revised]{12 March 2009}
\received[accepted]{5 June 2009}

\maketitle

\section{Introduction}

Industrial Control Systems (ICS) form the backbone of numerous critical infrastructures (CI), including the electric power grid and water treatment facilities. These systems manage the physical operations within a CI through the use of computing and communication components such as Programmable Logic Controllers (PLCs), Supervisory Control and Data Acquisition (SCADA) systems, and communication networks~\cite{ScanningTheCycle_AsiaCCS2021}. Although automation has streamlined the monitoring and control of these critical infrastructures, it has simultaneously made them vulnerable to potential threats from malicious actors, as evidenced by various attacks~\cite{weinbergerStuxnet,ukraineBlackout,germanSteelMill}. The increasing frequency of attacks on Industrial Control Systems (ICS) has spurred extensive research into security measures aimed at prevention, mitigation, and response~\cite{challenges_SnP_ahmed2020}. The prior research focus lies in two key areas: comprehensive testing of ICS security and the development of robust intrusion detection techniques. The success of these initiatives largely hinges on the system's design and the availability of data especially the attack examples, obtained from critical infrastructures. 



In this work, we present a novel technique for the automatic generation of attacks based on both data and design principles, aimed at driving an Industrial Control System (ICS) into an anomalous state. This resource is invaluable for encompassing a wide range of attack and anomaly scenarios. We employ agents powered by Large Language Models (LLMs) to automatically identify attack patterns from historical ICS data. Traditionally, manually crafted attacks~\cite{sridhar_compsac_2016} depend on human expertise to induce an anomalous state. By leveraging the pattern extraction capabilities of LLMs, we can analyze expert-developed action sets designed to compromise system safety, thereby uncovering novel attack sequences from the data. This approach allows us to identify previously unseen attacks and assess their impact, with the goal of exploring numerous possibilities for disrupting a physical process and pinpointing combinations of sensors and actuators that can be manipulated to achieve this.

We apply our proposed approach to a scaled-down version of a water treatment plant, known as the SWaT testbed~\cite{mathurTippenhauer}, as a case study to generate attack patterns. Our automated method produced a greater number of attack patterns compared to those generated manually by human experts. In contrast, experts familiar with the SWaT testbed~\cite{mathurTippenhauer} designed a set of 36 attack scenarios across the plant by operating the system for five days~\cite{sridhar_dataset_paper}. This achievement is significant given the limited research facilities and the scarcity of attack data. However, for machine learning algorithms that require large datasets, the available data is insufficient for training a robust supervised learning attack detection model. We validated the attack patterns generated by \emph{AttackLLM} using three methods: (i) validating the normal patterns generated by \emph{AttackLLM} and comparing them with the design specifications, (ii) comparing the attacks generated by \emph{AttackLLM} with those created by human experts, and (iii) comparing the results between two different LLM Agents. These automatically generated attacks can serve as valuable resources for enhancing our understanding of potential threats and for developing effective attack detection strategies.

\begin{figure}
    \centering
    \includegraphics[width=\linewidth]{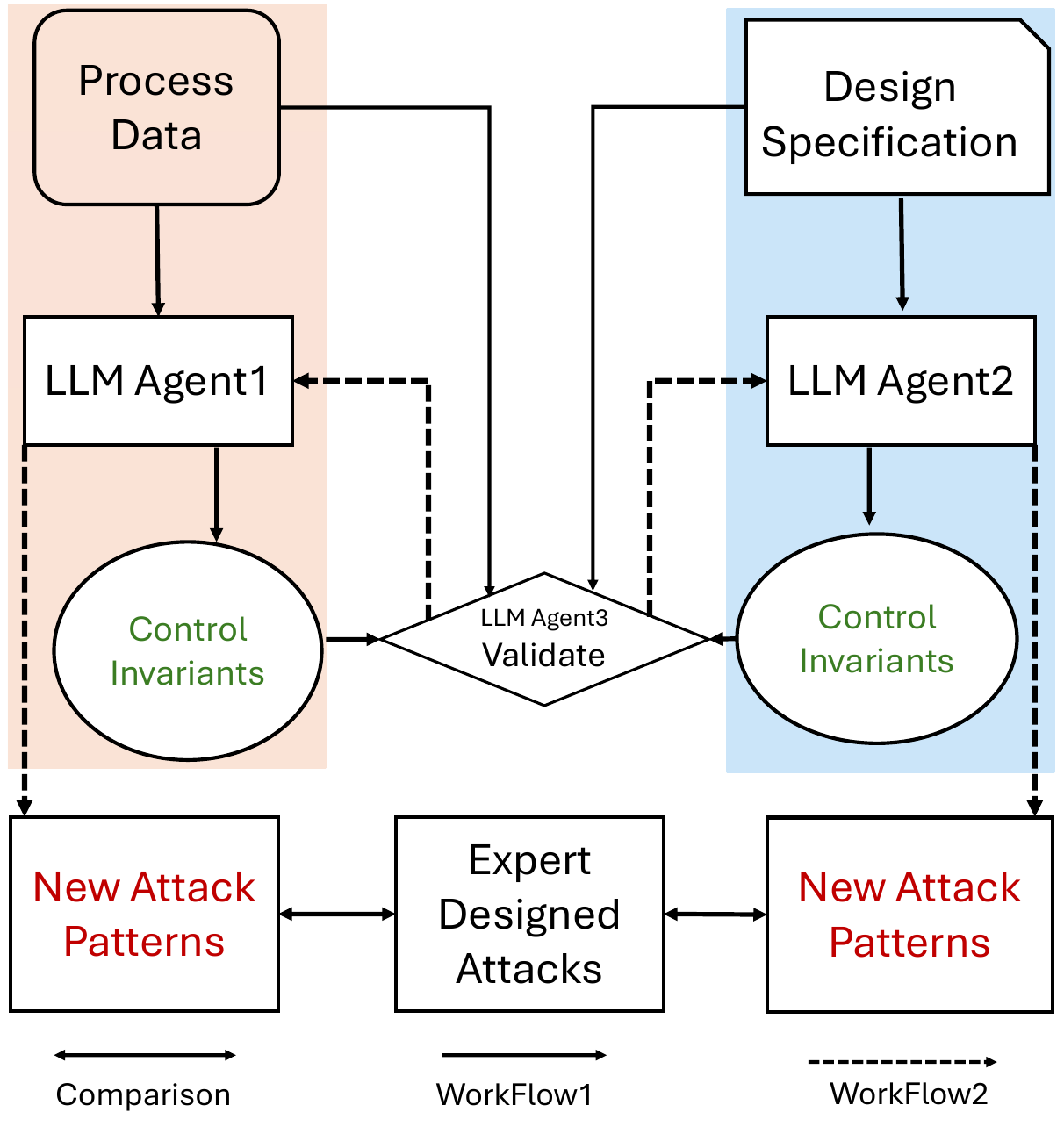}
    \caption{Overview of the workflows. Two LLM models are used to obtain the control invariants. An LLM based agent validates the control invariants. Workflow2 presented as dashed lines, take the validated control invariants and outputs new attack patterns, which in turn are analysed in comparison with an human expert designed attacks.}
    \label{fig:overview}
\end{figure}

\section{Background and System Description}

\subsection{Large Language Models}

Modern advancements in natural language processing (NLP) are driven by large language models (LLMs), including architectures like GPT-4~\cite{Brown2020} and Llama~\cite{Touvron2023}. These models leverage extensive pre-training on diverse textual datasets to develop sophisticated language comprehension and generative capabilities, enabling tasks such as context-aware reasoning and human-like text production. 
Modern LLM-driven systems~\cite{Muthusamy2023,Xi2025,Wang2023} integrate large language models to analyze problems, formulate actionable strategies, and deploy solutions through tool-assisted execution. For example, LLM-based infrastructure management systems can dynamically analyze server logs, diagnose anomalies, and autonomously execute corrective actions or escalate alerts to human operators. 
Within industrial control systems (ICS) and anomaly detection, we investigate the feasibility of using LLMs to autonomously derive physical invariants from operational documentation and sensor data. This approach aims to (1) replace labor-intensive manual extraction processes and (2) uncover intricate or non-obvious invariants that human analysts might overlook. By harnessing the contextual reasoning of LLMs, we seek to advance the scalability and precision of attack generation frameworks for securing cyber-physical infrastructure.




\subsection{System Overview}

The Figure~\ref{fig:overview} illustrates a systematic workflow for analyzing the control system of a water treatment plant, generating control invariants, and discovering new attack patterns. The process begins with two primary inputs: the dataset, which contains time-series data from the plant's sensors and actuators (e.g., water level, valve status, and pump status), and the {design documentation}, which provides contextual information about the system's layout, sensor-actuator relationships, and operational constraints. These inputs are independently processed by two LLM agents: {LLM Agent 1} analyzes the dataset to infer control invariants, while {LLM Agent 2} extracts control invariants from the design documentation. The generated invariants are then validated by {LLM Agent 3}, which ensures their consistency with the observed behavior in the dataset. Once validated, the invariants are used by LLM Agent 1 and LLM Agent 2 to generate potential attack patterns that exploit vulnerabilities in the system. These LLM-generated attacks are compared with attacks manually generated by a {domain expert} to evaluate their quality and coverage. The comparison identifies gaps and discrepancies, leading to the discovery of {new attack patterns} that are added to the knowledge base. This workflow combines data-driven analysis, design insights, and expert input to enhance the system's security by identifying previously unknown vulnerabilities and improving its resilience against potential threats.


\subsection{SWaT Testbed and Dataset Description}

The Secure Water Treatment (SWaT) testbed, located at the Singapore University of Technology and Design (SUTD), is a fully operational water treatment facility designed to replicate real-world industrial processes. This testbed has been widely adopted by researchers as a benchmark for evaluating defense mechanisms in Industrial Control Systems (ICS)~\cite{mathurTippenhauer, swatDataset}. The SWaT system is capable of producing 5 gallons of treated water per minute, employing a multi-stage process that includes ultrafiltration, reverse osmosis, and chemical dosing. The SWaT testbed consists of six distinct stages, each equipped with a variety of sensors and actuators. These sensors monitor critical water parameters, such as tank levels, flow rates, pressure, pH, oxidation-reduction potential, and conductivity. Actuators, including motorized valves and electric pumps, are used to control the treatment process. Communication within the system is facilitated through two networks: a Level 0 network connects sensors and actuators to Programmable Logic Controllers (PLCs), while a Level 1 network enables inter-PLC communication.

\subsection*{Dataset Description}
The dataset used in this study is sourced from the historian server of the SWaT testbed. It includes readings from all sensors and actuators, collected at a sampling rate of one reading per second. This dataset is publicly available and is particularly valuable for ICS research as it contains both normal operational data and attack scenarios introduced by human experts. The attacks are carefully designed to mimic real-world threats, making the dataset a robust benchmark for evaluating the effectiveness of anomaly detection and attack generation methods~\cite{sridhar_dataset_paper}. The SWaT dataset was chosen for two primary reasons:
\begin{enumerate}
    \item \textbf{Wide Adoption in ICS Research}: The dataset is extensively used in the ICS community, providing a common ground for comparing results across studies.
    \item \textbf{Expert-Designed Attacks}: The inclusion of attacks crafted by human experts ensures that the dataset reflects realistic threat scenarios, making it an ideal benchmark for evaluating the quality of synthetically generated attacks.
\end{enumerate}



\subsection{Invariants in Industrial Control Systems}

Invariants represent physical conditions or relationships among process variables that must always hold true for a system to operate under normal conditions. These invariants are rooted in the physical laws governing the system's processes~\cite{Choi_ControlInvariants_CCS2018}. For instance, in a water treatment system, the relationship between the water level in a tank and the flow rates through the inlet and outlet valves must conform to specific physical laws, as measured by sensors. Violations of these invariant rules serve as strong indicators of anomalies, which may arise due to faults or malicious attacks, that's why we use control invariants as an inpit to LLMAgents to generate attack examples. Traditionally, invariant rules are manually defined by system engineers during the design phase of Industrial Control Systems (ICS). However, this manual approach is often time-consuming, prone to errors, and may fail to capture all relevant invariants, particularly in complex systems~\cite{feng2019systematic_Venkat_NDSS2019}.


 Control invariants  model both control and physical properties/states of the physical process. The control invariants are determined jointly by the control algorithm, and the laws of physics. These control invariants reflect (and set constraints on) an plant’s normal behaviors according to its control inputs and current physical states. The state-space representation of the system is given by the following equations:

\begin{equation}
    \mathbf{x}(k+1) = \mathbf{A}\mathbf{x}(k) + \mathbf{B}\mathbf{u}(k) + \mathbf{v}(k)
\end{equation}

\begin{equation}
    \mathbf{y}(k) = \mathbf{C}\mathbf{x}(k) + \boldsymbol{\eta}(k)
\end{equation}

\noindent where: $\mathbf{x}(k)$ is the state vector at time step $k$, $\mathbf{u}(k)$ is the input vector at time step $k$, $\mathbf{y}(k)$ is the output vector at time step $k$, $\mathbf{A}$, $\mathbf{B}$, and $\mathbf{C}$ are system matrices, $\mathbf{v}(k)$ is the process noise, and $\boldsymbol{\eta}(k)$ is the measurement noise. Based on data and design, LLMAgents extracted the following code snippet (see Listing~\ref{lst:swatControlLogic}) demonstrating the control logic for SWaT Stage 1, including the motorized valve, pumps, and flow meter:

\begin{lstlisting}[style=structuredText, caption={Control Logic for SWaT Stage 1}, label={lst:swatControlLogic}]
(* Control Logic for SWaT Stage 1 *)

(* Motorized Valve MV101 *)
IF LIT101 < 250 THEN
    Alarm := TRUE;
    P101 := 1; (* STOP *)
    P102 := 1; (* STOP *)
ELSIF LIT101 > 1200 THEN
    Alarm := TRUE;
ELSIF LIT101 < 500 THEN
    MV101 := 2; (* OPEN *)
ELSIF LIT101 > 800 THEN
    MV101 := 1; (* CLOSE *)
END_IF;

(* Pump P1 and P2 *)
IF LIT301 < 800 THEN
    P101 := 2; (* START *)
    P102 := 2; (* START *)
ELSIF LIT301 > 1000 THEN
    P101 := 1; (* STOP *)
    P102 := 1; (* STOP *)
END_IF;

(* Flowmeter FIT201 *)
IF FIT201 < 0.5 THEN
    P101 := 1; (* STOP *)
    P102 := 1; (* STOP *)
END_IF;
\end{lstlisting}


\begin{table*}[h!]
\centering
\caption{Control Invariants for Stage1.}
\label{tab:control_invariants_deepseek}
\begin{tabular}{|c|p{8cm}|p{7cm}|}
\hline
\textbf{Invariant} & \textbf{Description} & \textbf{Condition} \\ \hline
1 & MV101 opens when LIT101 < 500mm & \texttt{IF LIT101 < 500mm THEN MV101 = OPEN} \\ \hline
2 & MV101 closes when LIT101 > 800mm & \texttt{IF LIT101 > 800mm THEN MV101 = CLOSE} \\ \hline
3 & Alarm and pump stop when LIT101 < 250mm & \texttt{IF LIT101 < 250mm THEN ALARM \& P101/P102 = STOP} \\ \hline
4 & Alarm when LIT101 > 1000mm & \texttt{IF LIT101 > 1000mm THEN ALARM} \\ \hline
5 & P101/P102 starts when LIT301 < 800mm & \texttt{IF LIT301 < 800mm THEN P101/P102 = START} \\ \hline
6 & P101/P102 stops when LIT301 > 1000mm & \texttt{IF LIT301 > 1000mm THEN P101/P102 = STOP} \\ \hline
7 & P101/P102 stops when FIT201 < 0.5 m³/h & \texttt{IF FIT201 < 0.5 m³/h THEN P101/P102 = STOP} \\ \hline
8 & LIT101 must remain between 250mm and 1000mm & \texttt{250mm $\leq$ LIT101 $\leq$ 1000mm} \\ \hline
9 & LIT301 must remain between 800mm and 1000mm during normal operation & \texttt{800mm $\leq$ LIT301 $\leq$ 1000mm} \\ \hline
10 & FIT201 must remain $\geq$ 0.5 m³/h when pumps are running & \texttt{IF P101/P102 = START THEN FIT201 $\geq$ 0.5 m³/h} \\ \hline
11 & P102 starts if P101 fails or conditions (LIT301 < 800mm or FIT201 < 0.5 m³/h) & \texttt{IF P101 = FAIL OR (LIT301 < 800mm OR FIT201 < 0.5 m³/h) THEN P102 = START} \\ \hline
\end{tabular}
\end{table*}

\begin{table*}[ht]
\centering
\begin{tabularx}{\textwidth}{lXX} 
\toprule
Invariant & Validation Result & Reason \\
\midrule
IF LIT101 < 250 THEN Alarm \& P101/P102 STOP & Not Passed & No instances found where LIT101 < 250 and both P101 and P102 are stopped \\
IF LIT101 > 1200 THEN Alarm & Not Passed & No instances found where LIT101 > 1200 \\
IF LIT101 < 500 THEN MV101 = 2 (OPEN) & Passed &  \\
IF LIT101 > 800 THEN MV101 = 1 (CLOSE) & Passed &  \\
IF LIT301 < 800 THEN P101/P102 = 2 (START) & Passed &  \\
IF LIT301 > 1000 THEN P101/P102 = 1 (STOP) & Not Passed & No instances found where LIT201 > 1000 and both P101 and P102 are stopped \\
IF FIT201 < 0.5 THEN P101/P102 = 1 (STOP) & Passed &  \\
\bottomrule
\end{tabularx}
\caption{Control Invaiants Validation. Common reason for not validating a control is due to absence of control example in data.}
\label{tab:control_invariant_validation_copilot}
\end{table*}







\begin{figure}
    \centering
    \includegraphics[width=\linewidth]{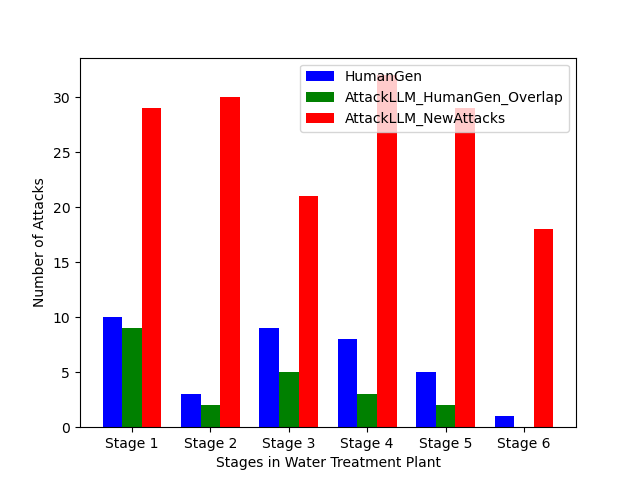}
    \caption{Comparison between human-generated attacks from~\cite{sridhar_dataset_paper}, and \emph{AttackLLM} generated attacks. Besides new attacks, we see that there is a huge overlap between human and \emph{AttackLLM} attacks.}
    \label{fig:comparison_figure}
\end{figure}

\subsection{Physical Invariant Inference and Validation}

In the initial phase, large language models (LLMs) are employed to infer potential physical invariants by analyzing the dataset and testbed specification documents. Microsoft Copilot which is based on the GPT-4 architecture and DeepSeek-V3 are used to extract control invariants and generate attack patterns.  In the second phase, the control invariants generated in the first phase are rigorously validated. The underlying principle is that a valid physical invariant should remain consistent throughout the data collection period. Specifically, the parameters derived from the invariant relationships should exhibit stability when calculated using different segments of the training dataset. If these parameters fluctuate significantly over short intervals, the invariant may lead to inaccurate representation of the physical process. To validate the invariants, we employ a multi-fold cross-validation approach. The training dataset is divided into multiple segments, and the parameters are derived from each segment. These parameters are then applied to the remaining segments to check for consistency. A control invariant is confirmed only if the derived parameters remain stable across all segments.

\section{Results}


We have formulated the following research questions to be answered by this work: 


\noindent \textbf{RQ1: Can \emph{AttackLLM} autonomously generate and empirically validate control invariants?}
To investigate this, we analyze AttackLLM’s ability to derive control invariants using SWaT (Stage 1) as a case study. Listing~\ref{lst:swatControlLogic} illustrates the extracted control logic from both the operational dataset and system documentation, highlighting actuator-driven water flow regulation and sensor-triggered alarms. Table~\ref{tab:control_invariants_deepseek} enumerates invariants generated by \emph{AttackLLM}, which successfully capture inter-stage dependencies and sensor-actuator relationships. Validation results (Table~\ref{tab:control_invariant_validation_copilot}) indicate that the majority of invariants demonstrate validity when tested against operational datasets. However, failure cases arose when invariants referenced states were not found in normal operational data (e.g., Row 1 in Table~\ref{tab:control_invariant_validation_copilot}), suggesting dataset limitations rather than model inaccuracies. \emph{AttackLLM} exhibits high accuracy in logical reasoning, achieving exceptional precision in inferring invariants for complex systems such as water treatment processes.

\noindent \textbf{RQ2: Can AttackLLM autonomously generate novel attack patterns?}
Table~\ref{tab:attack_patterns_stage1_deepseek} and Table~\ref{tab:attack_patterns_copilot_stage1} present attack patterns generated by DeepSeek and Copilot, respectively. Analysis reveals three key insights between the models. DeepSeek: (Table~\ref{tab:attack_patterns_stage1_deepseek}) uncovered an unexpected inter-stage attack targeting sensor LIT301 in Stage~3, alongside a multi-stage attack (\emph{A26}), which aligns with human expert-generated attacks in prior work \cite{sridhar_dataset_paper}. Copilot: (Table~\ref{tab:attack_patterns_copilot_stage1}) exclusively generated a distinct multi-stage attack (\emph{A21}), with all other attacks being consistent across both models. These results demonstrate \emph{AttackLLM’s} capability to identify both known and novel attack vectors, including cross-stage dependencies, in complex systems.

\noindent \textbf{RQ3: Can AttackLLM autonomously generate stealthy attack vectors?}
Table~\ref{tab:stealthy_attack_patterns_stage1_deepseek} and Table~\ref{tab:stealthy_attack_patterns_stage1_copilot} summarize stealthy attacks generated by DeepSeek and Copilot, respectively. Both models successfully generated stealthy attacks, including incremental sensor manipulations (e.g., Row~1 in Table~\ref{tab:stealthy_attack_patterns_stage1_deepseek} proposes a 1mm perturbation per time instance). Notably, this pattern aligns with the human expert-generated attack \emph{A3} documented in \cite{sridhar_dataset_paper}, demonstrating AttackLLM’s ability to replicate known adversarial strategies. Two major classes of attacks emerge in all six stages, 1) gradual change attack, and 2) intermittent disruption attack. 

\noindent \textbf{RQ4: How do \emph{AttackLLM}-generated attacks compare to human expert-derived attack patterns?}
A comparative analysis was conducted against the human expert-generated attacks in \cite{sridhar_dataset_paper}. As shown in Figure~\ref{fig:comparison_figure}, \emph{AttackLLM} successfully replicated nine out of ten human-designed attacks, including complex multi-stage scenarios, while generating 20 novel attack patterns for Stage~1. The most significant advancement lies in stealthy attack generation, where \emph{AttackLLM} proposed innovative strategies to evade detection mechanisms—a critical contribution to understanding adversarial resilience in cyber-physical systems.

Figure~\ref{fig:comparison_figure} summarizes attack patterns across all six stages of the water treatment process. \emph{AttackLLM} generated 159 total attacks, with 120 empirically validated as legitimate—a multifold increase over the 36 human-generated attacks documented in \cite{sridhar_dataset_paper}. Due to space limitations, only the attack tables for Stage~1 are included here, whereas Figure~\ref{fig:comparison_figure} summarizes the results for all the six stages in the SWaT testbed. All attack patterns and an extended version of this paper will be made available on a dedicated website upon acceptance.

\section{Related Work}
The generation of malicious examples for evaluating the robustness of machine learning algorithms in Industrial Control Systems (ICS) has been a topic of significant interest in recent years. Traditional approaches rely heavily on domain expertise and the availability of testbeds to collect both normal and attack data \cite{sridhar_dataset_paper}. However, these methods are often limited by the high cost of human expertise and the scarcity of accessible testbeds, making the process inefficient and impractical for large-scale applications \cite{Adepu2016}. Existing datasets for ICS environments, such as those from the SWaT and WADI testbeds \cite{Mathur2016}, have been widely used for anomaly detection and attack pattern analysis. While these datasets provide valuable insights, they are often constrained by the limited scope of attack scenarios and the reliance on pre-existing attack examples \cite{Urbina2016}. This limitation hinders the development of robust anomaly detection methods, as the diversity and complexity of attack patterns are often insufficient to fully evaluate the resilience of machine learning models \cite{Lin2018}. Some other efforts, such as those on the EPIC testbed (smart grid testbed), highlight the challenges of collecting attack data due to safety concerns. For instance, it was not even possible to collect attack data or examples in the EPIC testbed because of the risks associated with compromising a live smart grid system \cite{Ahmed2021}. This further underscores the need for synthetic data generation methods that do not rely on physical testbeds.

Recent advancements in large language models (LLMs) and generative AI have opened new avenues for generating synthetic data and attack patterns. LLMs have demonstrated remarkable capabilities in understanding and generating complex patterns, making them suitable for applications in cybersecurity \cite{Brown2020}. 
However, their application in ICS environments remains underexplored. The integration of LLMs with data-centric and design-centric methodologies, as proposed in this work, represents a novel approach to addressing the challenges of attack pattern generation in ICS.

A recent survey by \cite{ammara2025syntheticnetworktrafficdata} provides a comprehensive comparison of synthetic network traffic data generation techniques, highlighting the strengths and limitations of various approaches. This study emphasizes the importance of high-quality synthetic data for training and evaluating machine learning models in cybersecurity applications. Previous studies have a major focus on Generative Adversarial Network (GAN)-based methods. While GANs have been widely used to generate adversarial examples against specific intrusion detection systems (IDS), these examples are often tailored to particular systems and lack generalizability across the problem domain. Additionally, the survey highlights that much of the attention in previous studies has been directed toward network-layer traffic and IoT datasets, with limited exploration of synthetic data generation for Industrial Control Systems (ICS). This gap underscores the need for domain-specific approaches, such as the one proposed in our work, which leverages advanced generative techniques to create realistic and generalizable attack patterns tailored to ICS environments.

Our work builds on these foundations by leveraging LLMs to generate high-quality attack patterns that surpass those created by human experts. By combining data-driven analysis with design insights, our approach eliminates the need for expensive testbeds and pre-existing attack examples, offering a scalable and cost-effective solution for enhancing the security of ICS environments. This multi-agent framework represents a significant advancement in the field, providing a promising avenue for improving the robustness and resilience of machine learning algorithms in critical infrastructure systems.

\section{Conclusions}
This work demonstrates that large language models (LLMs) can autonomously generate and validate attack patterns for industrial control systems (ICS) by synthesizing process data, system documentation, and control logic. Through the development of \emph{AttackLLM}, we show that LLM-driven agents can: \noindent \textbf{1) Derive and Validate Control Invariants}: \emph{AttackLLM} successfully inferred sensor-actuator relationships and inter-stage dependencies in the SWaT testbed, achieving high validation accuracy (RQ1). While dataset limitations occasionally restricted invariant verification, the model demonstrated robust logical reasoning capabilities for complex systems. \noindent \textbf{2) Generate Novel and Stealthy Attack Vectors}: Beyond replicating human expert-designed attacks (e.g., \emph{A3}, \emph{A26}), \emph{AttackLLM} generated 20 novel attack patterns for Stage~1 and identified 159 total attacks across all six stages—a multi-fold increase over prior human efforts (RQ2–4). Notably, it uncovered stealthy strategies such as incremental perturbations (1mm/time instance) and intermittent disruptions, which evade conventional detection mechanisms.

\noindent \textbf{Enhance Adversarial Understanding}: By autonomously discovering cross-stage attack dependencies (e.g., targeting LIT301 in Stage~3 from Stage~1), \emph{AttackLLM} provides insights into systemic vulnerabilities that human analysts might overlook.

\noindent \textbf{Implications for ICS Security}: The scalability and precision of \emph{AttackLLM} highlight its potential as a tool for proactive defense, enabling rapid stress-testing of anomaly detection systems and the identification of previously unknown attack surfaces.

\noindent \textbf{Limitations and Future Work}: While our results are promising, reliance on operational datasets restricted invariant validation in edge-case scenarios. Future work will extend \emph{AttackLLM} to other ICS domains (e.g., power grids) and integrate adversarial training to refine detection evasion strategies. Additionally, we will explore hybrid human-AI frameworks to address dataset gaps and improve interpretability.

\begin{table*}[h!]
\centering
\begin{tabular}{|p{2cm}|p{2cm}|p{6cm}|p{6cm}|}
\hline
\textbf{Attack}/(Ref~\cite{sridhar_dataset_paper}) & \textbf{Target} & \textbf{Manipulated Value} & \textbf{Impact} \\ \hline
LIT101 (A33,A36) & Sensor & Set LIT101 > 1000mm or LIT101 < 250mm & Tank overflow or underfill, pump stoppage \\ \hline
MV101/(A1) & Actuator & Force MV101 = 2 or MV101 = 1 & Uncontrolled inflow or no inflow \\ \hline
P101/P102 (A2,A34,A35) & Actuator & Force P101/P102 = 2 or P101/P102 = 1 & Over-pumping or no pumping \\ \hline
LIT301 & Sensor & Set LIT301 < 800mm or LIT301 > 1000mm & Incorrect pump operation \\ \hline
 FIT201 & Sensor & Set FIT201 < 0.5 m³/h & Pump stoppage, dry running \\ \hline
Simultaneous Sensor Manipulation/(A26) & Multiple Sensors & Set LIT101 > 1000mm, LIT301 < 800mm, FIT201 < 0.5 m³/h & Cascading failure, tank overflow, pump stoppage \\ \hline
Physical Tampering & Sensors/Actuators & Damage or disable components & Unpredictable system behavior \\ \hline
Denial of Service (DoS) & Communication Network & Flood control system with traffic & Uncontrolled system operation \\ \hline
\end{tabular}
\caption{Attack Patterns generated by DeepSeek for SWaT Satge~1.}
\label{tab:attack_patterns_stage1_deepseek}
\end{table*}







\begin{table*}[h!]
\centering
\begin{tabular}{|p{3cm}|p{2cm}|p{6cm}|p{5cm}|}
\hline
\textbf{Attack/(Ref~\cite{sridhar_dataset_paper})} & \textbf{Target} & \textbf{Manipulated Value} & \textbf{Impact} \\ \hline
Gradual Drift in LIT101/(A3) & Sensor (LIT101) & Slowly increase or decrease LIT101 readings by a small margin (e.g., 1mm/min). & Causes gradual tank overflow or underfill without triggering alarms. \\ \hline
Intermittent MV101 Jamming & Actuator (MV101) & Randomly force MV101 to open/close for short durations, mimicking normal operation. & Disrupts water inflow subtly, leading to inconsistent tank levels over time. \\ \hline
Pump Efficiency Degradation & Actuator (P101/P102) & Gradually reduce pump efficiency (e.g., reduce flow rate by 1\% every hour). & Causes slow depletion or overfilling of the tank, avoiding immediate detection. \\ \hline
FIT201 Flow Rate Spoofing & Sensor (FIT201) & Spoof FIT201 readings to show a slightly lower flow rate (e.g., 0.6 m³/h instead of 0.7 m³/h). & Causes pumps to operate inefficiently, leading to energy waste or dry running over time. \\ \hline
LIT301 Level Spoofing & Sensor (LIT301) & Spoof LIT301 readings to show a slightly lower level (e.g., 790mm instead of 800mm). & Causes pumps to start prematurely, leading to over-pumping and energy waste. \\ \hline
Time-Delayed Sensor Data & Multiple Sensors & Introduce a small delay (e.g., 1-2 seconds) in sensor data transmission. & Causes control logic to operate on outdated data, leading to inefficiencies or instability. \\ \hline
Intermittent Sensor Noise & Multiple Sensors & Add random noise to sensor readings within acceptable limits. & Causes control system to make suboptimal decisions, leading to gradual system degradation. \\ \hline
Selective Data Suppression & Communication Network & Suppress specific sensor data packets intermittently (e.g., LIT101 data every 5 minutes). & Causes control system to miss critical data, leading to inefficiencies or instability. \\ \hline
\end{tabular}
\caption{Stealthy Attack Patterns generated by DeepSeek for SWaT Stage~1.}
\label{tab:stealthy_attack_patterns_stage1_deepseek}
\end{table*}

\bibliographystyle{ACM-Reference-Format}
\bibliography{sample-base,references}

\newpage
\appendix

\section{Tables for Attack Patterns}

\begin{table*}[h!]
\centering
\begin{tabular}{|>{\raggedright}p{3cm}|>{\raggedright}p{6cm}|>{\raggedright\arraybackslash}p{6cm}|}
\hline
\textbf{Attack/(Ref~\cite{sridhar_dataset_paper})} & \textbf{Values of Sensors and Actuators} & \textbf{Impact} \\
\hline

\textbf{Spoofing LIT101 Readings (A33)} & 
Spoof LIT101 readings to show a water level below 250mm or above 1000mm. \newline
\textbf{Duration:} Until the system triggers an alarm or stops the pumps. & 
Causes the system to trigger alarms and potentially stop the pumps, leading to damage or overfilling. \\
\hline

\textbf{Disabling MV101 (A1)} & 
Disable MV101 to prevent it from opening or closing. \newline
\textbf{Duration:} Until the system detects the malfunction. & 
Causes uncontrolled water inflow, leading to overfilling or depletion of the tank. \\
\hline


\textbf{Spoofing FIT201 Readings} & 
Spoof FIT201 readings to show a flow rate below 0.5 m³/h. \newline
\textbf{Duration:} Until the system triggers an alarm or stops the pumps. & 
Causes the system to stop the pumps, leading to dry running or insufficient water flow. \\
\hline

\textbf{Force On Outlet Pump -- P101} & 
\textbf{P101}: Force to stay on (P101 = 2) regardless of control signals. \newline
\textbf{Duration}: Until the tank is drained or the attack is detected. \newline
\textbf{LIT101}: Normal operation. \newline
\textbf{MV101}: Normal operation. & 
\textbf{Dry Run}: The tank will be drained, potentially causing the pump to run dry, leading to pump damage or insufficient water supply. \\
\hline

\textbf{Force Off Outlet Pump -- P101/(A34,A35)} & 
\textbf{P101}: Force to stay off (P101 = 1) regardless of control signals. \newline
\textbf{Duration}: Until the tank overflows or the attack is detected. \newline
\textbf{LIT101}: Normal operation. \newline
\textbf{MV101}: Normal operation. & 
\textbf{Overflow}: The tank will continue to fill with water, leading to an overflow and potential water damage or system shutdown. \\
\hline

\textbf{Combined Attack: False Low Level + Force Open Inlet Valve/(A21)} & 
\textbf{LIT101}: Set to a value below the lower threshold (e.g., 110) continuously. \newline
\textbf{MV101}: Force to stay open (MV101 = 2) regardless of control signals. \newline
\textbf{P101}: The control system will turn off the pump (P101 = 1). & 
\textbf{Rapid Overflow}: The tank will quickly fill with water, leading to an overflow and potential water damage or system shutdown. \\
\hline

\end{tabular}
\caption{Attack Patterns generated by Copilot for SWaT Stage~1.}
\label{tab:attack_patterns_copilot_stage1}
\end{table*}


\begin{table*}[h!]
\centering
\begin{tabular}{|p{3cm}|p{7cm}|p{5cm}|}
\hline
\textbf{Attack} & \textbf{Values of Sensors and Actuators} & \textbf{Impact} \\
 \hline

\textbf{Gradual False Low Level Reading} & 
\textbf{LIT101}: Gradually decrease the reading to a value below the lower threshold (e.g., 110) over a longer period. \newline
\textbf{Duration}: Over several hours or days. \newline
\textbf{MV101}: The control system will gradually open the inlet valve (MV101 = 2). \newline
\textbf{P101}: The control system will gradually turn off the pump (P101 = 1). & 
\textbf{Overflow}: The tank will eventually overflow, but the gradual change will make it less suspicious and delay detection. \\
\hline

\textbf{Gradual False High Level Reading} & 
\textbf{LIT101}: Gradually increase the reading to a value above the upper threshold (e.g., 140) over a longer period. \newline
\textbf{Duration}: Over several hours or days. \newline
\textbf{MV101}: The control system will gradually close the inlet valve (MV101 = 1). \newline
\textbf{P101}: The control system will gradually turn on the pump (P101 = 2). & 
\textbf{Dry Run}: The tank will eventually be drained, but the gradual change will make it less suspicious and delay detection. \\
\hline

\textbf{Intermittent Force Open Inlet Valve (MV101)} & 
\textbf{MV101}: Intermittently force to stay open (MV101 = 2) for short periods. \newline
\textbf{Duration}: Random intervals over several hours or days. \newline
\textbf{LIT101}: Normal operation. \newline
\textbf{P101}: Normal operation. & 
\textbf{Overflow}: The tank will eventually overflow, but the intermittent changes will make it less suspicious and delay detection. \\
\hline

\textbf{Intermittent Force Closed Inlet Valve (MV101)} & 
\textbf{MV101}: Intermittently force to stay closed (MV101 = 1) for short periods. \newline
\textbf{Duration}: Random intervals over several hours or days. \newline
\textbf{LIT101}: Normal operation. \newline
\textbf{P101}: Normal operation. & 
\textbf{Insufficient Water Supply}: The tank will eventually be drained, but the intermittent changes will make it less suspicious and delay detection. \\
\hline

\textbf{Intermittent Force On Outlet Pump (P101)} & 
\textbf{P101}: Intermittently force to stay on (P101 = 2) for short periods. \newline
\textbf{Duration}: Random intervals over several hours or days. \newline
\textbf{LIT101}: Normal operation. \newline
\textbf{MV101}: Normal operation. & 
\textbf{Dry Run}: The tank will eventually be drained, but the intermittent changes will make it less suspicious and delay detection. \\
\hline

\textbf{Intermittent Force Off Outlet Pump (P101)} & 
\textbf{P101}: Intermittently force to stay off (P101 = 1) for short periods. \newline
\textbf{Duration}: Random intervals over several hours or days. \newline
\textbf{LIT101}: Normal operation. \newline
\textbf{MV101}: Normal operation. & 
\textbf{Overflow}: The tank will eventually overflow, but the intermittent changes will make it less suspicious and delay detection. \\
\hline

\textbf{Combined Stealthy Attack: Gradual False Low Level + Intermittent Force Open Inlet Valve} & 
\textbf{LIT101}: Gradually decrease the reading to a value below the lower threshold (e.g., 110) over a longer period. \newline
\textbf{MV101}: Intermittently force to stay open (MV101 = 2) for short periods. \newline
\textbf{P101}: The control system will gradually turn off the pump (P101 = 1). & 
\textbf{Rapid Overflow}: The tank will eventually overflow, but the combined gradual and intermittent changes will make it less suspicious and delay detection. \\
\hline

\end{tabular}
\caption{Stealthy Attack Patterns generated by Copilot for SWaT Stage~1.}
\label{tab:stealthy_attack_patterns_stage1_copilot}
\end{table*}

\end{document}